\documentclass[conference]{IEEEtran}
\usepackage{balance}
\usepackage[a4paper, left=1.27cm, right=1.27cm, top=1.815cm, bottom=4.3cm]{geometry}

\usepackage{float}
\usepackage{standard}
\usepackage{etex}

\usepackage{rotate}
\usepackage{algorithmic}
\usepackage{algorithm}
\usepackage{float}
\usepackage{tikz}
\usepackage{refcount}

\usepackage{pgf,tikz}
\usepackage{pgfplots}
\usepackage{pgfplotstable}
\usepackage{bm}
\usepackage{adjustbox}
\usetikzlibrary{calc}
\usepackage{relsize}
\usepackage{ifthen}

\usetikzlibrary{calc,fit,arrows,plotmarks,intersections,patterns,shapes,decorations,positioning, backgrounds, matrix}

\usepackage[numbers,sort&compress]{natbib}
\usepackage[ngerman, english]{babel}

\usetikzlibrary{arrows.meta}
\edef\wdArrowLength{2}
\tikzset{>={Latex[width=1.5mm,length=\wdArrowLength mm]}}
\usepackage{graphicx}

\usepackage{cancel}
\usepackage{mathtools}
\usepackage{shadethm}
\usepackage{empheq}
\usepackage{skull}
\usepgfplotslibrary{units}
\usepackage{calc}
\usepackage{nicefrac}
\usepackage{fancybox}
\usepackage{psfrag}
\usepackage{dsfont}

\title{Accelerated Recovery with RIS: Designing Wireless
Resilience in Mission-Critical Environments}
\author{\IEEEauthorblockN{Kevin Weinberger$^{\ast }$, Robert-Jeron Reifert$^{\ast }$, Aydin Sezgin$^{\ast }$ and Mehdi Bennis$^{\dagger }$}
\IEEEauthorblockA{\IEEEauthorrefmark{1}Institute of Digital Communication Systems, Ruhr University Bochum, Germany\\
\IEEEauthorrefmark{2}Centre for Wireless Communications, University of Oulu, Finland \\
Email: {\{kevin.weinberger,robert-.reifert,aydin.sezgin\}@rub.de}, mehdi.bennis@oulu.fi}
\thanks{This work was supported in part by the German Research Foundation (DFG) in the course of the project SPP2433 under the project no. 541021107 (Measurement Technology on Flying Platforms) under grant SE 1697/22-1. This work was supported in part by the German Federal Ministry of
Education and Research (BMBF) in the course of the 6GEM Research Hub
under grant 16KISK037.}
}
\date{\today}

\usepackage{graphicx}
\usepackage{placeins}
\usepackage{booktabs}
\usepackage{marvosym}
\usepackage{multirow}

\usepackage{latexsym, amsmath, amssymb, amsfonts, upgreek}
\usepackage[nolist]{acronym}

\usepackage{tikz}
\usetikzlibrary{calc,arrows,positioning,decorations,shadows,shapes,fadings,matrix}
\tikzset{>=latex'}
\tikzset{semithick}

\makeatletter
\providecommand{\IfElsePackageLoaded}[3]{\@ifpackageloaded{#1}{#2}{#3}}
\makeatother

\usepackage{subfigure}
\IfElsePackageLoaded{subfig}
	{\usepackage[subfigure]{tocloft}}{	
	\IfElsePackageLoaded{subfigure}
		{\usepackage[subfigure]{tocloft}}
		{\usepackage{tocloft}}
	}

\makeatletter

\def\tikz@delimiter#1#2#3#4#5#6#7#8{%
	\bgroup
		\pgfextra{\let\tikz@save@last@fig@name=\tikz@last@fig@name}%
		node[outer sep=0pt,inner sep=0pt,draw=none,fill=none,anchor=#1,at=(\tikz@last@fig@name.#2),#3]
		{%
			{\nullfont\pgf@process{\pgfpointdiff{\pgfpointanchor{\tikz@last@fig@name}{#4}}{\pgfpointanchor{\tikz@last@fig@name}{#5}}}}%
			\delimitershortfall\z@% as suggested by morbusg (maximum space not covered by a delimiter = 0)
			\resizebox*{!}{#8}{$\left#6\vcenter{\hrule height .5#8 depth .5#8 width0pt}\right#7$}%
		}
		\pgfextra{\global\let\tikz@last@fig@name=\tikz@save@last@fig@name}%
	\egroup%
}

\tikzset{hexagon/.code={
	\draw (0,2) -- (-4,0) -- (0,-2) -- (4,0) -- (0,2);
}}

\tikzset{phone/.code={
   \node [rectangle,rounded corners=1.5pt,draw,minimum height=0.6cm, minimum width=0.35cm] at (0,0){};
   \node [rectangle,rounded corners=1.5pt,draw,minimum height=0.5cm, minimum width=0.3cm] at (0,0){};
}}

\makeatletter
\def\cantox@vector#1#2#3#4#5#6#7#8{%
  \dimen@.5\p@
  \setbox\z@\vbox{\boxmaxdepth.5\p@
   \hbox{\kern-1.2\p@\kern#1\dimen@$#7{#8}\m@th$}}%
  \ifx\canto@fil\hidewidth  \wd\z@\z@ \else \kern-#6\unitlength \fi
  \ooalign{%
    \canto@fil$\m@th \CancelColor
    \vcenter{\hbox{\dimen@#6\unitlength \kern\dimen@
      \multiply\dimen@#4\divide\dimen@#3 \vrule\@depth\dimen@\@width\z@
      \vector(#3,-#4){#5}%
    }}_{\raise-#2\dimen@\copy\z@\kern-\scriptspace}$%
    \canto@fil \cr
    \hfil \box\@tempboxa \kern\wd\z@ \hfil \cr}}
\def\bcancelto#1#2{\let\canto@vector\cantox@vector\cancelto{#1}{#2}}
\makeatother

\IEEEoverridecommandlockouts
\usepackage{algorithmic}
\begin{document}
\maketitle
\begin{abstract}
With the advent of 6G and beyond, connectivity is no longer just about linking devices, it is defined as the network’s ability to maintain communication under varying link conditions. This evolution makes wireless networks the backbone of critical operations, where resilience becomes more essential than ever. Beyond supporting traditional services, 6G also enables innovative applications that were previously impossible, opening new opportunities for next-generation wireless systems. As a result, there is a pressing demand for strategies that can adapt to dynamic channel conditions, interference, and unforeseen disruptions, ensuring seamless and reliable performance in an increasingly complex environment. Despite considerable research, existing resilience assessments lack comprehensive key performance indicators (KPIs), especially those quantifying its adaptability, which are vital for identifying a system's capacity to rapidly adapt and reallocate resources. In this work, we bridge this gap by proposing a novel framework that explicitly quantifies the adaption performance by augmenting the gradient of the system's rate function. To further enhance the network resilience, we integrate Reconfigurable Intelligent Surfaces (RISs) into our framework due to their capability to dynamically reshape the propagation environment while providing alternative channel paths. Numerical results show that gradient augmentation enhances resilience by improving adaptability under adverse conditions while proactively preparing for future disruptions.
\end{abstract} 
%\begin{IEEEkeywords}
%
%%resilience, reconfigurable intelligent surface (ris), intelligent reflecting surface (irs), cloud radio access network (c-ran) resource allocation, quality of service.
%\end{IEEEkeywords}
\thispagestyle{empty}
\pagestyle{empty}

\section{Introduction}
%mixed criticality, resilience, RIS.}\\
%\cite{8493070}: Closed-form blockage probability of open park-like scenario.\\
%\cite{9834416}: \emph{Proactive Resilience in 1-2-1 Networks}, mmWave network, link layer perspective from Information Theory\\
%\cite{9112354}: mmWave channel in details, beamforming in details, some kind of resilience in their \emph{anti-blocker} algorithm\\
%\cite{9847340}: Strong reference, we can closely follow some of their parts. E.g., imperfect CSI, statisitcal CSI, IRS model, LoS channel model, SINR closed-form expression, channel estimation error, alternating optimization (receiver filter, power allocation, IRS phase shifts) \\
Future 6G networks are being shaped by recent advancements and global research, aiming to support both human-centric and \ac{AI}-driven applications with seamless connectivity \cite{you2021towards}.
This evolution envisions the realization of emerging applications such as autonomous transportation, remote-controlled robotics, and smart power grids. In these cases, even a brief network failure could lead to severe consequences, from safety hazards to operational breakdowns.
It follows, that the network must not just support
connectivity but also exhibit adaptive, context-aware, and autonomous
behaviors.
Interestingly, despite these challenges, much of the current research on 5G and 6G has focused primarily on reliability and robustness \cite{reliability,Robustness}, with less attention given to overall resilience, the ability to adapt and recover quickly from failures \cite{resiliencemetric}.
In fact, resilience encompasses a multitude of aspects, by demanding the network to not only absorb disturbances and ensure operational stability but also to integrate adaptive mechanisms that enable autonomous recovery, or even thrive \cite{AntifragileRS}, from unexpected challenges \cite{RobertRes,brosInArms},

Despite extensive research efforts, current resilience assessments in wireless networks still lack comprehensive \acp{KPI} \cite{ResByDesign}. Particularly those that quantify adaptability, which is critical for evaluating a system’s ability to rapidly adapt and efficiently reallocate resources in changing conditions. To bridge this gap, we propose a novel framework that explicitly quantifies adaptability by augmenting the gradient of the system's rate function. In this work, the rate gradient refers to the derivative of the achievable rate with respect to system parameters, used to adapt transmission rates dynamically. The augmented gradient quantifies network responsiveness, enabling a balanced trade-off between robustness and adaptability. Our unified approach ensures stability during prolonged disruptions while swiftly adapting to sudden changes, enhancing service continuity and quality even when suffering from unexpected failures.

%To implement our approach, it becomes necessary to not only to design the allocation of the network's resources but also to actively modify the radio channel.
To implement our approach, resource allocation in the network must be supported by the active modification of radio channels.
To this end, we integrate \acp{RIS} into our design \cite{basar2019wireless, SynBenefits}. With their ability to dynamically reconfigure the wireless propagation environment and establish alternative channel paths, \acp{RIS} are shown to significantly enhance network resilience \cite{RIS_RES,CognitiveResilience}. In the following sections, we detail our methodology and present extensive numerical results that validate the effectiveness of our approach.

\section{System Model}\label{ch:Sysmod}
\begin{figure}
	\centering \includegraphics[width=1\linewidth]{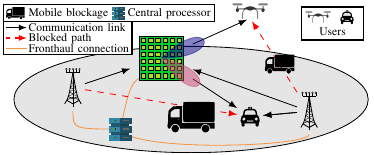}%2.75
	\caption{System model with mission-critical communications where mobile blockages can block the direct AP-user paths within the cohererence interval.}
	\label{fig:1}
\end{figure}%

This paper considers the \ac{RIS}-aided cell-free \ac{MIMO} downlink system depicted in Fig.~\ref{fig:1}. In details, a set of single-antenna users $\mathcal{K}=\{1,\cdots,K\}$ is being served by a set of $L$-antenna \acp{AP} $\mathcal{N}=\{1,\cdots,N\}$. To enable an extra set of channel-links, we deploy an $M$-element \ac{RIS}, configured as a uniform planar array\footnote{The elements are arranged in a grid that is as close to square as possible while maintaining integer dimensions.}, within user range (i.e., no farther than 300m from the users).  The position is chosen in a strategic way, so that the \ac{RIS} is capable of providing alternative paths should direct links become blocked. The \ac{RIS} and the \acp{AP} are connected to and managed by a \ac{CP} via orthogonal fronthaul links. To meet the demands of the users, a \ac{QoS} target is set as a desired data rate denoted as $r_k^\mathsf{des}$, for each user.

\subsection{Channel Model}\label{ssec:chan}
The channel model in this paper assumes quasi-static block fading, where channel coefficients remain constant within the coherence time $T_\mathsf{c}$ but vary independently across coherence blocks. The direct channel links between \ac{AP} $n$ and user $k$ are subject to Rayleigh fading and denoted as $\vect{h}_{n,k} \in \mathbb{C}^{L\times1}$. The RIS-assisted reflected channels are modeled as \ac{LoS} given by
$
\mat{G}_{n,k} = \mat{H}_{n} \mathsf{diag}(\vect{v})\vect{g}_k \in \mathbb{C}^{L \times 1},
$
where $\mat{H}_{n} \in \mathbb{C}^{L\times M}$ represents the AP-to-RIS link, $\vect{g}_k \in \mathbb{C}^{M\times 1}$ the RIS-to-user link, and $\mathsf{diag}(\vect{v})\in \mathbb{C}^{M \times M}$ the reflection coefficient matrix, with $\vect{v} = [v_1,v_2,\dots v_M] \in \mathbb{C}^{M \times 1}$. Each reflection coefficient is defined as $v_m = e^{j\theta_m}$, where $\theta_m \in [0,2\pi]$ is the phase shift at the $m$-th RIS element.

The aggregate direct channel vector for user $k$ is given by
$
\vect{h}_k= [\vect{h}_{1,k}^T,\vect{h}_{2,k}^T,\dots,\vect{h}_{N,k}^T]^T \in \mathbb{C}^{NL\times 1},
$
while the combined AP-to-RIS channel matrix is
$
\vect{H}=[\vect{H}_{1}^T,\vect{H}_{2}^T,\dots,\vect{H}_{N}^T]^T \in \mathbb{C}^{NL \times M}.
$
Similarly, the aggregate transmit signal vector is
$
\vect{x}=[\vect{x}_1^T,\vect{x}_2^T,\dots, \vect{x}_N^T]^T \in \mathbb{C}^{NL \times 1}.
$

Using these aggregate representations, the received signal at user $k$ is expressed as the sum of the direct and reflected channel components:
\begin{align}\label{recSgn}
  y_k =& (\vect{h}_k + \mat{G}_k\vect{v})^H \vect{x} + n_k = (\vect{h}_k^\mathsf{eff})^H \vect{x} + n_k,
\end{align}
where $\mat{G}_k=\mat{H}\text{diag}(\vect{g}_k)$, $\vect{h}^\mathsf{eff}_k = \vect{h}_k + \mat{G}_k\vect{v}$ and $n_k\sim\mathcal{C}\mathcal{N}(0,\sigma_k)$ represents the additive white Gaussian noise (AWGN).

The symbols intended to be decoded by user $k$ are denoted by $s_k$. We assume these messages form an \ac{i.i.d.} Gaussian codebook. The symbols for user $k$ are transmitted by the n-th \ac{AP} using the beamforming vector $\vect{w}_{n,k} \in \mathbb{C}^{L\times 1}$, both provided by the \ac{CP}.

\newcommand{\RisChanFull}[1]{\vect{h}_{#1} + \mat{G}_{#1}\vect{v}}
\newcommand{\RisChan}[1]{\vect{h}^\mathsf{eff}_{#1}}

Hence, the overall transmit signal vector at the $n$-th \ac{AP} is given as
\begin{align}
\vect{x}_{n}=\sum_{k\in\mathcal{K}} \vect{w}_{n,k} s_k, \ \forall n \in \mathcal{N},
\end{align}
which is subject to the power constraint
$
\mathbb{E}\{\vect{x}_n^H \vect{x}_n\} \leq P^{\mathsf{max}}_n,
$
that can be expressed as
\begin{align}\label{eq:powConst}
\sum_{k\in\mathcal{K}}|| \vect{w}_{n,k}||_2^2 \leq P^{\mathsf{max}}_n , \ \forall n \in \mathcal{N}.
\end{align}

\noindent
The received signal (\ref{recSgn}) at user $k$ is then given by
\begin{align}
y_k = (\RisChan{k})^H \vect{w}_k s_k + \sum_{i\in\mathcal{K}\setminus\{k\}} (\RisChan{k}) \vect{w}_i s_i + n_k,
\end{align}
where the first term represents user $k$'s desired signal, and the second term accounts for all the other user's interference.

Thus, we formulate the \ac{SINR} of user $k$ decoding its message as
\begin{align}
\Gamma_k = \frac{|(\RisChan{k})^H \vect{w}_k|^2}{\sum_{i\in\mathcal{K}\setminus\{k\}}|(\RisChan{k})^H \vect{w}_i|^2 + \sigma^2},
\end{align}
where $\sigma^2$ denotes the noise power.

Using these definitions, each user's \ac{QoS} demands is satisfied, if the following condition holds:
\begin{align}
r_k^\mathsf{des} \leq r_k \leq B \log_2(1+\Gamma_k) ,
\end{align}
where $B$ denotes the transmission bandwidth and $r_k$ represents the allocated rate of user $k$.
\newcommand{\RisOnlyChan}[1]{\vect{h}_{#1}^\mathsf{RIS}}

When the direct links to user $k$ are blocked, by denoting $\RisOnlyChan{k} = \mat{G}_k\vect{v}$, we can define the RIS-link \ac{SINR} $\Gamma_k^\mathsf{RIS}$ and rate expressions $r_k^\mathsf{RIS}$ as
\begin{align}
&\Gamma_k^\mathsf{RIS} = \frac{|(\RisOnlyChan{k})^H \vect{w}_k|^2}{\sum_{i\in\mathcal{K}\setminus\{k\}}|(\RisOnlyChan{k})^H \vect{w}_i|^2 + \sigma^2}, \\
&r_k^\mathsf{RIS} \leq B \log_2(1+\Gamma_k^\mathsf{RIS}),\label{RIS}
\end{align}
which effectively represents the case, in which user $k$ is only served over the RIS-assisted links. Since we quantify the network's adaptability with the rate's gradient, by denoting $a_{k,i}=\vect{v}^H \mat{G}_k^H \vect{w}_i$ and $\vect{b}_{k,i} = \vect{w}_i^H \mat{G}_k\vect{v}\mat{G}_k^H\vect{w}_i\in \mathbb{C}^{M\times 1}$ we can express the gradient of the RIS-link rate \ac{w.r.t.} the phase shifts as
\begin{align}\label{gradV}
\nabla_\vect{v}r_k^\mathsf{RIS} =  \frac{2B(\vect{b}_{k,k} - \Gamma_k^\mathsf{RIS} \vect{b}_{k,i})}{\ln(2)(\sum_{i\in \mathcal{K}}{|a_{k,i}|^2} + \sigma^2)}.
\end{align}
\textit{Proof:} For a detailed derivation, we refer to Appendix \ref{sec:AppA} 
\section{RIS in the Context of Resilience}
\subsection{Resilience Metric}
In the context of resilience, wireless networks must maintain reliable performance even under dynamic conditions, such as link blockages or resource fluctuations. To ensure this, the network operates with a predefined target throughput dictated by the \ac{QoS} requirements, where $r_k^\mathsf{des}$ represents user $k$'s desired \ac{QoS} level. These \ac{QoS} demands remain constant within the coherence interval, while the network’s sum throughput varies based on allocated resources at time $t$, expressed as $\sum_{k=1}^{K} r_k(t)$, where $r_k(t)$ denotes the allocated data rate for user $k$ at time $t$. To characterize resilience behavior, we define two key moments in time: $t_0$, denoting the initial time when degradation manifests, and $t_{q}$ denoting the time at which the system reaches a recovered state. Building on the resilience metric proposed in \cite{RobertRes}, we adopt the network’s absorption, adaptation, and time-to-recovery metrics as cornerstones of the resilience performance. More precisely, the absorption metric is given by
\begin{align}\label{eq:res1}
r_\mathsf{abs} = \frac{1}{K} \sum_{k \in \mathcal{K}} \frac{r_k(t_0)}{r_k^\mathsf{des}},
\end{align}
which quantifies the system’s ability to sustain performance when facing disruptions. The adaptation metric, expressed as
\begin{align}\label{eq:res2}
r_\mathsf{ada} = \frac{1}{K} \sum_{k \in \mathcal{K}} \frac{r_k(t_{q})}{r_k^\mathsf{des}},
\end{align}
captures how well the network recovers. Finally, the time-to-recovery metric measures the speed of resilience and is given by
\begin{align}\label{eq:res3}
r_\mathsf{rec} =
\begin{cases}
1, & \text{if } t_{q} - t_0 \leq T_0 \\
\frac{T_0}{t_{q} - t_0}, & \text{otherwise}.
\end{cases}
\end{align}
where, $T_0$ represents the desired recovery time, defining the duration for which a functionality degradation is considered tolerable. A linear combination of the equations (\ref{eq:res1})-(\ref{eq:res3}) defines the considered resilience metric as
\begin{align}\label{eq:r}
r = \lambda_1 r_\mathsf{abs} + \lambda_2 r_\mathsf{ada} + \lambda_3 r_\mathsf{rec},
\end{align}
where the fixed weights $\lambda_i$, for $i \in \{1, 2, 3\}$, reflect the network operator’s priorities, such as emphasizing robustness, adaptation quality, or recovery time. These non-negative weights satisfy $\sum_{i=1}^{3} \lambda_i = 1$, ensuring that the best-case resilience value is $r = 1$.

\subsection{RIS as Resiliency Mechanism}
As previously discussed, an effective resiliency mechanism should promptly and efficiently address disruptions. However, outages arise for various reasons, each with different severities and impacts on the system. In this work, we propose leveraging the \ac{RIS} to enhance resilience. More precisely, integrating the \ac{RIS} introduces alternative reconfigurable channel paths to each user. Even when direct paths from all \acp{AP} to a user are entirely obstructed, the system can still recover by leveraging \ac{RIS}-assisted paths. In fact, without these alternative paths, communication with a blocked user could fail entirely in the presence of disruptions beyond the system's capacity to mitigate. In terms of the resilience metric, the inherent redundancy that the \ac{RIS} links contribute to the system can also be utilized to increase robustness \cite{RISCoverage}. Additionally, the reconfigurability of the phase shifts at the \ac{RIS} provides the system with an extra degree of freedom, which enables more effective rerouting of the signal in case of blockages. This translates to the \ac{RIS} aiding in adapting quicker to disturbances, improving adaption performance $r_\mathsf{ada}$, while also decreasing the time-to-recovery $r_\mathsf{rec}$ \cite{RIS_RES}.
This becomes an essential characteristic, especially when the resilience actions must restore system performance within a short desired recovery time \(T_0\). To this end, we propose to quantify this property by introducing the rate gradient \( \nabla_\vect{v} r_k^\mathsf{RIS} \) as a key metric. Maximizing its norm enhances user rate sensitivity, allowing rapid adaptation to disruptions like total \ac{LoS} loss. As a result, the \ac{RIS} enables balancing of both, redundancy and adaptability by providing redundant paths and reconfigurable phase shifts, respectively.

\section{Problem Formulation}
To analyze the impact of the \ac{RIS} on resilience performance, we formulate a multi-objective optimization problem that provides precise control over the individual components of the resilience metric. Furthermore, we propose to optimize this problem not only \textit{during} the recovery phase, but also \textit{before} any blockage occurs. The objective is to enable enhanced recovery from disturbances while simultaneously fortifying the system against future failures. The inclusion of multiple objectives is motivated by the versatility introduced through the weight parameters $\lambda_i$ in (\ref{eq:r}), which determine the relative emphasis placed on each aspect of the metric. First, we intend to capture the adaption metric $r_\mathsf{ada}$ by considering the network wide adaption gap \cite{RobertRes,RIS_RES} $\Uppsi =  \sum_{k\in \mathcal{K}} \big|\frac{r_k}{r_k^\mathsf{des}}-1\big|$, which represents the overall goal of serving each user with its required \ac{QoS}. Second, we employ gradient augmentation to enchance adaption performance, which boosts both, the time-to-recovery $r_\mathsf{rec}$ as well as the adaption $r_\mathsf{ada}$. This is achieved by maximizing the norm of the gradient of the RIS-assisted rate \ac{w.r.t.} the phase shift vectors $||\nabla_\vect{v} r_k^\mathsf{RIS}||_2$. Third, we provide robustness by including a RIS-link-redundancy term, represented by the difference between the total rate and the RIS-only rate as $\Delta_k^\mathsf{RIS} = |r_k-r^\mathsf{RIS}_k|^2$. By reducing this term we facilitate that some portion of the received signal at user $k$ is routed through the \ac{RIS}-assisted links.
With the above considerations, the overall problem can be formulated as
	\begin{align}\label{Prob1}
		\underset{\vect{w},\vect{v},\vect{r}}{\min} \quad & \nu_\mathsf{const}\Uppsi -  \sum_{k\in\mathcal{K}}\alpha_{1,k}||\nabla_\vect{v} r_k^\mathsf{RIS}||_2+ \sum_{k\in\mathcal{K}}\alpha_{2,k} \Delta_k  \tag{P1} \\
        \text{s.t.}\quad & (\ref{eq:powConst}), \nonumber
        \end{align}\vspace{-0.85cm}
        \begin{align}
        r_k &\leq B \log_2(1+\Gamma_k) , \, &&\forall k \in \mathcal{K}, \label{rateConst}\\
        r_k^\mathsf{RIS} &\leq B \log_2(1+\Gamma_k^\mathsf{RIS}) , \, &&\forall k \in \mathcal{K}, \label{rateConst2}\\
        |v_m| &= 1 ,\ &&\forall m \in \{1,\dots,M\}, \label{unit_mod}
	\end{align}
where $\vect{r} = [r_1, r_2, \dots, r_K, r_1^\mathsf{RIS}, r_2^\mathsf{RIS}, \dots, r_K^\mathsf{RIS}]^T$ represents the stacked rate vector, and the unit modulus constraints in (\ref{unit_mod}) enforce the phase shift conditions $0\leq \theta_m \leq 2\pi, \forall m \in \{1,\dots,M\}$. The weight $\nu_\mathsf{const}$ is set to a very high value to ensure that the adaptation gap $\Uppsi$ always takes priority, while $\alpha_{1,k}$ and $\alpha_{2,k}$ enable versatility in designing the adaptability-robustness trade-off for each user individually. Additionally, this approach enables efficient resource allocation so that users that are mission-critical or in vulnerable states become more resilient against failures.

Solving problem (\ref{Prob1}) is complex due to the interdependence between the variables $\vect{w}$ and $\vect{v}$ in $\nabla_\vect{v}r_k^\mathsf{RIS}$, as well as the constraints (\ref{rateConst}) and (\ref{rateConst2}). Further minimizing the norm of the gradient and satisfying the unit modulus constraint are additional challenges, especially when coupled with the rate constraints in (\ref{rateConst}) and (\ref{rateConst2}). These elements together make the optimization problem non-trivial, requiring specialized techniques such as alternating optimization and \acp{SCA} to handle these interdependencies effectively \cite{SynBenefits}. In fact, both sub-problems can be efficiently solved within the same \ac{SCA} framework \cite{SynBenefits}. As a result, full convergence of one sub-problem is not required before moving to the other; instead, just one iteration of each sub-problem can be performed before switching to the other. This approach significantly accelerates the overall convergence process, which is crucial in resilience scenarios and allows the evaluation of the resilience performance without the need to converge \cite{RIS_RES}.
	\subsection{Beamforming Design}
	As a result of the alternating optimization approach, the phase shifters $\vect{v}$ are assumed to be fixed for the duration of the beamforming design. Thus, problem (\ref{Prob1}) can be written as
		\begin{align}\label{Prob2}
		\underset{\vect{w},\vect{r},\vect{q},\vect{u}}{\min} \quad & \nu_\mathsf{const}\Uppsi  - \sum_{k\in \mathcal{K}} \alpha_{1,k} u_k +
 \sum_{k\in \mathcal{K}} \alpha_{2,k} \Delta_k
\tag{P2} \\
		\text{s.t.}\quad & (\ref{eq:powConst}), \nonumber
\end{align}\vspace{-0.85cm}
\begin{align}
		 \,\,\ r_k &\leq B \log_2(1+q_k) , \, &&\forall k \in \mathcal{K}, \label{P2rate}\\
r_k^\mathsf{RIS} &\leq B \log_2(1+q_k^\mathsf{RIS}) , \, &&\forall k \in \mathcal{K}, \label{P2rate2}\\
		q_k  \,\,\ &\leq \Gamma_k , \, &&\forall k \in \mathcal{K}, \label{P2SINR}\\
	q_k^\mathsf{RIS} &\leq \Gamma_k^\mathsf{RIS} , \, &&\forall k \in \mathcal{K}, \label{P2SINR2}\\
u_k \,\,\  &\leq\sum_{k\in\mathcal{K}}||\nabla_\vect{v} r_k^\mathsf{RIS}||_2, \, &&\forall k \in \mathcal{K},\label{gradVP1}\\
		\:\,\:\,\vect{q} \:\,&\geq 0 , \,\, \vect{u} \:\,\geq 0 \label{P2t1}
	\end{align}
where the introduction of the slack variables $\vect{q}=[q_1 ,\dots ,q_K,q^\mathsf{RIS}_1 ,\dots ,q^\mathsf{RIS}_K]$ convexify the rate expressions for the effective and RIS-link and the slack variables $\vect{u}=[u_1 ,\dots ,u_K]$ linearize the norm in the objective function. Furthermore, (\ref{P2t1}) signifies that all values in $\vect{q}$, and also $\vect{u}$, are nonnegative. However, the constraints in (\ref{P2SINR})-(\ref{gradVP1}) remain non-convex but can be rendered in a convex form through the \ac{SCA} approach. To this end, we rewrite (\ref{P2SINR}) as
	\begin{align}\label{SINR_rewritten}
		\sum_{i\in\mathcal{K}\setminus\{k\}}|(\RisChan{k})^H \vect{w}_i|^2 + \sigma^2 - \frac{|(\RisChan{k})^H \vect{w}_k|^2}{q_k}\leq 0.
	\end{align}
By applying the first-order Taylor approximation around the point 	$(\tilde{\vect{w}},\tilde{\vect{q}})$  to the fractional term, we obtain the following convex approximation of (\ref{SINR_rewritten}) \cite{SynBenefits}
	\begin{align}\label{SINR_convex1}
	\sum_{i\in\mathcal{K}\setminus\{k\}}|(\RisChan{k})^H \vect{w}_i|^2 + \sigma^2 +
			\frac{|(\RisChan{k})^H \tilde{\vect{w}}_k|^2}{(\tilde{q}_k)^2}q_k \nonumber \\ - \frac{2 \text{Re} \{\tilde{\vect{w}}_k^H(\RisChan{k})(\RisChan{k})^H\vect{w}_k \}}{\tilde{q_k}} \leq 0, \forall k \in \mathcal{K}.
	\end{align}
Due to the same structure, the same process can be repeated for the RIS-link \ac{SINR} in (\ref{P2SINR2}), by substituting $\vect{h}_k^\mathsf{eff}$ with $\vect{h}_k^\mathsf{RIS}$ and $q_k$ with $q_k^\mathsf{RIS}$, respectively:
	\begin{align}\label{SINR_convex2}
	\sum_{i\in\mathcal{K}\setminus\{k\}}|(\RisOnlyChan{k})^H \vect{w}_i|^2 + \sigma^2 +
			\frac{|(\RisOnlyChan{k})^H \tilde{\vect{w}}_k|^2}{(\tilde{q}_k^\mathsf{RIS})^2}q_k \nonumber \\ - \frac{2 \text{Re} \{\tilde{\vect{w}}_k^H(\RisOnlyChan{k})(\RisOnlyChan{k})^H\vect{w}_k \}}{\tilde{q}_k^\mathsf{RIS}} \leq 0, \forall k \in \mathcal{K}.
	\end{align}
Utilizing the homogeneity property of norms, the constraint in (\ref{gradVP1}) can also be reformulated in a similar structure
\begin{align}\label{gradReform}
\ln(2)\Big(\sum_{i\in \mathcal{K}}{|a_{k,i}|^2} + \sigma^2\Big) - \frac{2B||\vect{b}_{k,k} - q_k^\mathsf{RIS} \hspace{-0.3cm} \underset{ i\in\mathcal{K}\setminus\{k\}}{\sum}\hspace{-0.3cm}\vect{b}_{k,i}||_2}{u_k}& \leq 0, \nonumber\\ \forall k \in \mathcal{K}&,
\end{align}
where the fraction can be linearized by a Taylor approximation around the point $(\tilde{\vect{w}},\tilde{\vect{q}},\tilde{\vect{u}})$ as
\begin{align}
T^w_{k} = \beta_k|| \vect{\eta}_k || &-  \frac{\beta_k}{\tilde{u}_k}|| \vect{\eta}_k || (u_k-\tilde{u}_k) +\nonumber \\ \frac{\beta_k}{||\vect{\eta}_k||}\big( -\Re& \{ \vect{\eta}^H \underset{i\in\mathcal{K}\setminus\{k\}}{\sum}(\vect{b}_{k,i})\}({q}^\mathsf{RIS}_k-\tilde{{q}}^\mathsf{RIS}_k)\nonumber\\
+  \Re&\{\vect{\eta}^H( \mat{\Omega}_{k,k}^a + \mat{\Omega}_{k,k}^b)(\vect{w}_k-\tilde{\vect{w}}_k)\} \nonumber \\
- \sum_{i\in\mathcal{K}\setminus\{k\}}\Re&\{\vect{\eta}^H( \tilde{{q}}^\mathsf{RIS}_k( \mat{\Omega}_{k,i}^a + \mat{\Omega}_{k,i}^b))(\vect{w}_i-\tilde{\vect{w}}_i)\}  \big),
\end{align}
where $\beta_k=\frac{2B}{\tilde{u}_k}, \vect{\eta}_k = \tilde{\vect{b}}_{k,k}- \tilde{q}_k \sum_{i\in\mathcal{K}\setminus\{k\}}\tilde{\vect{b}}_{k,i}$, $\mat{\Omega}_{k,i}^a = \vect{w}_i^H \mat{G}_k \vect{v} \mat{G}_k^H $ and $\mat{\Omega}_{k,i}^b =\mat{G}_k^H \vect{w}_i (\mat{G}_k \vect{v})^H $.
Thus, the approximation of problem (\ref{Prob2}) can be written as
\begin{align}\label{Prob2.1}
	\underset{\vect{w},\vect{r},\vect{q},\vect{u}}{\min} \quad & \nu_\mathsf{const}\Uppsi - \alpha_{1,k} u_k +
 \sum_{k\in \mathcal{K}} \alpha_{2,k} \Delta_k \tag{P2.1} \\
	\text{s.t.}\quad & (\ref{eq:powConst}),(\ref{P2rate}),(\ref{P2rate2}), (\ref{P2t1}),  (\ref{SINR_convex1}), (\ref{SINR_convex2}), \nonumber\\
&\ln(2)\Big(\sum_{i\in \mathcal{K}}{|a_{k,i}|^2} + \sigma^2\Big) - T_k^w\leq 0, \forall k \in \mathcal{K}.
\end{align}
Problem (\ref{Prob2.1}) is convex and can be solved iteratively using the \ac{SCA} method. More precisely, we denote $\mat{\Lambda}_z^w = [\vect{w}_z^T ,\vect{\kappa}_z^T ]^T$ as a vector stacking the optimization variables of the beamforming design problem at iteration $z$, where $\vect{\kappa}_z = [\vect{r}_z^T,\vect{q}_z^T,\vect{u}_z]^T$. Similarly $\hat{\mat{\Lambda}}_z^w  = [\hat{\vect{w}}_z^T ,\hat{\vect{\kappa}}_z^T ]^T$ and $\tilde{\mat{\Lambda}}_z^w  = [\tilde{\vect{w}}_z^T ,\tilde{\vect{\kappa}}_z^T ]^T$ define the optimal solutions and the point, around which the approximations are computed, respectively. With these expressions, given a point $\tilde{\mat{\Lambda}}_{z}^w $, an optimal solution $\hat{\mat{\Lambda}}_z^w $ can be obtained by solving problem (\ref{Prob2.1}).

\newcommand{\RisChanv}[2]{\tilde{h}_{#1,#2} + \tilde{\mat{G}}_{#1,#2}\vect{v}}

\subsection{Phase Shift Design}
While designing the phase shifters at the \ac{RIS}, the beamformers are assumed to be fixed due to the alternating optimization approach. With the goal of using a comparable problem structure as in (\ref{Prob2}), we denote $|(\RisChan{i})^H \vect{w}_k|^2$ $=$ $\tilde{h}_{i,k} + \tilde{\mat{G}}_{i,k} \vect{v}$, where $\tilde{h}_{i,k} = \vect{w}_k^H \vect{h}_i$ and $\tilde{\mat{G}}_{i,k} = \vect{w}_k^H \mat{G}_i$. With the above definitions, the \ac{SINR} constraints can be written similar to (\ref{SINR_rewritten}) as
\begin{align}\label{SINR_rewritten_v}
	&\sum_{i\in\mathcal{K}\setminus\{k\}}\hspace{-0.3cm}|\RisChanv{k}{i}|^2 + \sigma^2 - \frac{|(\RisChanv{k}{k})|^2}{q_k}\leq 0,\\
	&\sum_{i\in\mathcal{K}\setminus\{k\}}|\mat{G}_{k,i}\vect{v}|^2 + \sigma^2 - \frac{|(\mat{G}_{k,k}\vect{v})|^2}{q_k}\leq 0 ,\forall k \in \mathcal{K}.\label{SINR_rewritten_v2}
\end{align}
By applying the same procedure used to obtain equations (\ref{SINR_convex1}) and (\ref{SINR_convex2}), the Taylor approximations for the SINR constraints around the phase shift vector $\vect{v}$ instead of $\vect{w}$ can be derived  \cite{SynBenefits,RIS_RES}.
Regarding the gradient-norm constraint in (\ref{gradVP1}), we utilize the reformulation in (\ref{gradReform}) and calculate the linearization around the point $(\tilde{\vect{v}},\tilde{\vect{q}},\tilde{\vect{u}})$ as \vspace{-0.15cm}
\begin{align}\label{gradApprxV}
T^v_{k} = \beta_k|| \vect{\eta}_k || &-  \frac{\beta_k}{\tilde{u}_k}|| \vect{\eta}_k || (u_k-\tilde{u}_k) \nonumber \\ +\frac{\beta_k}{||\vect{\eta}_k||}\big( &-\Re \{ \vect{\eta}^H \underset{i\in\mathcal{K}\setminus\{k\}}{\sum}(\tilde{\vect{b}}_{k,i})\}({q}^\mathsf{RIS}_k-\tilde{{q}}^\mathsf{RIS}_k)\nonumber\\
+  \Re\{\vect{\eta}^H&( \mat{S}_{k,k}-q_k^\mathsf{RIS} \sum_{i\in\mathcal{K}\setminus\{k\}}\mat{S}_{k,i})(\vect{v}_k-\tilde{\vect{v}}_k)\} \big), \\[-20pt] \nonumber 
\end{align}
where $\mat{S}_{k,i}= 2\mat{G}^H_k \vect{w}_i \vect{w}_i^H \mat{G}_k $. For the unit-modulus constraint in (\ref{unit_mod}), we adopt the penalty method \cite{RIS_RES}. Hence, the weighted first-order Taylor approximation of the term $\sum_{m=1}^{M}(|v_m|^2 -1)$, which is given by $\Phi = \alpha_{v}\sum_{m=1}^{M}\text{Re}\{2\tilde{v}_m^*v_m-|\tilde{v}_m|^2\}$ is included as a penalty in the objective function, where $\alpha_{v}$ is a weighting factor.
At this point, the approximated optimization problem for the phase shift design can be formulated as\vspace{-0.1cm}
\begin{align}\label{Prob3}
	\underset{\vect{v},\vect{r},\vect{q},\vect{u}}{\min}  \,\,\, \,\,& \nu_\mathsf{const}\Uppsi + \Phi  - \alpha_{1,k} u_k +
 \sum_{k\in \mathcal{K}} \alpha_{2,k} \Delta_k \tag{P3} \\
	\text{s.t.} \quad &(\ref{P2rate}),(\ref{P2rate2}),(\ref{P2t1}),(\widetilde{\ref{SINR_rewritten_v}}) ,(\widetilde{\ref{SINR_rewritten_v2}}), \nonumber\\
&\ln(2)(\sum_{i\in \mathcal{K}}{|a_{k,i}|^2} + \sigma^2) - T_k^v\leq 0, \forall k \in \mathcal{K}. \\[-16pt] \nonumber 
\end{align}
where $(\widetilde{\ref{SINR_rewritten_v}})$,$(\widetilde{\ref{SINR_rewritten_v2}})$  are the first-order Taylor approximation of {(\ref{SINR_rewritten_v})} and {(\ref{SINR_rewritten_v2})} around the point $(\tilde{\vect{v}},\tilde{\vect{q}},\tilde{\vect{u}})$, respectively, which are omitted for page length constraints reasons.
Due to the similarity of the problem formulation and the utilization of the same \ac{SCA} framework, problem (\ref{Prob3}) can be solved by defining $\mat{\Lambda}_z^v = [\vect{v}_z^T, \vect{\kappa}_z^T]^T$ and following the same iterative procedure as for solving problem (\ref{Prob2.1}).

\subsection{Resilience-Guided Alternating Optimization}
In the context of resilience, our goal is to find a solution that meets a specific trade-off defined by the weights $\lambda_i, \forall i\in\{1,2,3\}$ in (\ref{eq:r}) quickly. Therefore, instead of fully optimizing each sub-problem until convergence, we perform just one iteration for each sub-problem before switching to the other \cite{RIS_RES}. Once the adaptation gap converges, the algorithm then enhances adaptability and robustness to prepare for future failures while keeping the gap minimized.
By choosing the weights $\vect{\alpha} = [\alpha_{1,1}, \dots, \alpha_{1,K},\alpha_{2,1} \dots \alpha_{2,K}]$ accordingly, the algorithm can be tailored to respond effectively to different values of $\lambda_i$. The detailed steps of the resilience-guided alternating optimization are presented in Algorithm \ref{alg}, where $T_\mathsf{calc}$ is the time needed to compute the solution of a sub-problem and $T_\mathsf{c}$ is the coherence time.
\vspace{0pt}
\begin{algorithm}[H]
\setlength{\topsep}{0pt}
\setlength{\parskip}{0pt}
\footnotesize
\caption{Resiliency-aware Alternating Optimization}\label{alg}
\begin{tikzpicture}[>={Latex[length=5pt]}, baseline=(current bounding box.north)]\label{alg:AltOpt}
	\footnotesize
	\tikzset{myRect/.style={draw,rectangle,minimum width=0.5cm, minimum height=0.25cm,align=center}}
	\tikzset{myDiam/.style={draw,diamond,aspect=1.5,minimum width=1.25cm, minimum height=0.9cm,align=center}}
	\tikzset{myArrow/.style={->,draw,line width=0.25m}}
	\tikzset{myDot/.style={draw,minimum size=0.1,circle,fill,scale=0.15}}
	
	\node[align=center](input) at (-1,-1.85){ Input:\\ $\tilde{\vect{w}},\tilde{\vect{v}},\tilde{\vect{\kappa}},$\\ $T_\mathsf{c}$,$\vect{\alpha}$};
	
	\node[myRect](initLam) at ($(input)+(0,2.45)$){Create\\$\tilde{\mat{\Lambda}}_z^o$};
	%\node (initLamTxt) at (initLam){};

	\node[myDiam] (o1) at (0,0){};
	\node[] (o1txt) at (o1){$o=w$};
	\node[myRect] (P2) at ($(o1)+(2.625,0.3)$){$\hat{\vect{\Lambda}}_{z+1}^w \leftarrow$ solve (P2.1)};
	\node[myRect] (P3) at ($(o1)+(2.625,-0.3)$){$\hat{\vect{\Lambda}}_{z+1}^v \leftarrow$ solve (P3)};
	\node[myDiam] (oT) at (6.,-0.75){};
	\node[] (oTText) at (oT){$T{<}\:T_\mathsf{c}$};
	\node[myDiam] (o2) at (4.555,-1.5){};
	\node[] (o2txt) at (o2){$o=w$};
	\node[myRect,align=center] (Lambw) at ($(o2)+(-2.45,0.45)$){$[\tilde{\vect{w}}^T, \tilde{\vect{\kappa}}^T]^T \leftarrow \hat{\mat{\Lambda}}_z^w$ ,\\$\tilde{\mat{\Lambda}}_z^v \leftarrow [\tilde{\vect{v}}^T, \tilde{\vect{\kappa}}^T]^T $};
	\node[myRect,align=center] (Lambv) at ($(o2)+(-2.45,-0.45)$){$[\tilde{\vect{v}}^T, \tilde{\vect{\kappa}}^T]^T \leftarrow \hat{\mat{\Lambda}}_z^v$ \\ $\tilde{\mat{\Lambda}}_z^w \leftarrow [\tilde{\vect{w}}^T, \tilde{\vect{\kappa}}^T]^T$};
	
%	\node[myDiam](Psi) at ($(oT)+(-0.45,-1.025)$){};
%	\node(Psitxt) at (Psi) {$\Uppsi<\tau$};

	\draw[->] (input.north) |- ($(input.north)!0.25!(initLam.240)$) -|node[pos=0.76,anchor=210]{$\,\,\,\phantom{.}_{o\leftarrow w}$}node[pos=0.535,anchor=210]{$\phantom{.}_{T{\leftarrow 0}}$}node[pos=0.655,anchor=210]{$\phantom{.}_{z{\leftarrow 0}}$} (initLam.240);
	\draw[->] (initLam.south) |- (o1.west);

	\draw[->] (o1.east) --node[myDot,pos=1]{} ($(o1.east)+(0.1,0)$) |-node[pos=0.65,above]{\tiny$\mathsf{T}$} (P2.west) ;
	\draw[->] (o1.east) -- ($(o1.east)+(0.1,0)$)  |-node[pos=0.65,above]{\tiny$\mathsf{F}$} (P3.west) ;
	
	\node[myDot] (incrementDot) at ($(oT.north)-(1.9,-0.25)$){};
	\node[myDot] (incrementDot2) at ($(oT.north)-(1.75,-0.25)$){};
    \draw     (incrementDot) -- (incrementDot2);

    \draw[->] (P2.east) -| (incrementDot);
	\draw[->] (P3.east) -| (incrementDot);
	
	\draw[->] (incrementDot2) |-node[pos=0.775,below]{$\phantom{.}_{z\,{\leftarrow}{z+1}}$}node[pos=0.775,above]{ \tiny${T{\leftarrow}{T\hspace{-0.075cm}+\hspace{-0.075cm}T_\mathsf{calc}}}$} (oT.west);
\draw[->] (incrementDot2) |-node[pos=1,anchor=180]{Output:\\$\hat{\vect{w}}_z,\hat{\vect{v}}_z$}($(incrementDot2)+(0.5,0.5)$) ;
	
	%\draw[->] (oT.north) |- node[pos=0.1,right]{\tiny$\mathsf{T}$} node[pos=0.65,below]{$\tilde{\mat{\Lambda}}_z^o \leftarrow \hat{\mat{\Lambda}}_z^o$}  ($($(oT.north)!0.5!(o1.north)$) + (0,0.4) $) -| (o1.north);
	
	\draw[->] (oT.south) |- node[pos=0.35,anchor=-35]{\tiny$\mathsf{F}$}  (o2.east); %($(oT.west)!0.51!(o2.north)$)
	
	\draw[->] (o2.west) --node[myDot,pos=1]{} ($(o2.west)-(0.125,0)$) |-node[pos=0.65,above]{\tiny$\mathsf{T}$} (Lambw.east) ;
	\draw[->] (o2.west) -- ($(o2.west)-(0.125,0)$)  |-node[pos=0.65,above]{\tiny$\mathsf{F}$} (Lambv.east) ;
	
	\node[myDot] (reassignDot) at  ($(o2)-(4.39,0)$) {};
	\draw[->] (Lambw.west) -|node[pos=0.25,above]{$\phantom{.}_{o\leftarrow v}$} (reassignDot) ;
    \draw[->] (Lambv.west) -|node[pos=0.25,above]{$\phantom{.}_{o\leftarrow w}$} (reassignDot) ;

    \draw[->] (reassignDot) -|node[pos=0.65,anchor=-25]{} (o1.south);
	
	\path (oT.east) |-node[pos=0.85,anchor=125,align=center](a){} node[pos=0.15,anchor=35,align=center](b){\tiny$\mathsf{T}$} ($(oT.east)-(0.0,1)$);% -- (Psi.east); %(oT.east)!0.5!
	
\draw[->] (oT.east) -- ($(oT.east)-(0.0,1)$);
	\node[align=center] at ($(a)+(-0.1,-0.15)$){Stop};
	
%Output:\\$\hat{\vect{w}}_z,\hat{\vect{v}}_z$
%	\draw[->] (Psi.south) --node[pos=0.25,right]{\tiny$\mathsf{T}$} ($(Psi.south)+(0,-0.4)$);
	
%	\node at ($(Psi.south)+(0,-0.65)$) {Stop};
	
\end{tikzpicture}
\end{algorithm}

\section{Numerical results}
In this section, we numerically evaluate the performance of our proposed gradient augmentation approach in the context of resilience. To this end we assume a cell-free \ac{MIMO} system with $N=2$ \ac{AP}, each of which equipped with $L=8$ antennas and positioned on opposite corners of the area, which spans $[-500,500] \times [-500,500] \text{m}^2$. We assume $K=6$ single-antenna users to be distributed equally on a circle with a radius of 250m around the RIS, which is centrally positioned and consists of \( M = 500 \) reflective elements arranged in a quadratic grid with \( \lambda_f/4 \) spacing, where \( \lambda_f = 0.1 \)m is the wavelength. For the RIS channels, we use the correlated channel model from \cite{corrBj}. Further, reflected channels follow a line-of-sight model, while direct channels experience Rayleigh fading with log-normal shadowing (8dB standard deviation). Moreover, we assume a bandwidth of $B=10 \mathsf{MHz}$, a noise power of $\sigma^2$ = -100dBm, a maximum transmit power of $P_n^\mathsf{max} = 32$dBm, $T_0=0.15$s and each user to request a \ac{QoS} of $r_k^\mathsf{des}=6$ Mbps. It should be noted that our model assumes perfect \ac{CSI} and continuous user activity. In practice, channel estimation errors and bursty traffic patterns may occur, potentially reducing the achievable gains. A detailed treatment of these aspects is left for future work.

We define an outage as the event where each direct link between \ac{AP} $n$ and user $k$ is subject to a blockage, removing it from the network. RIS-assisted links are exempt, as the \ac{RIS} is positioned to bypass potential blockages, ensuring an unobstructed communication path. Additionally, blockages are considered from the strongest channel onward to force reoptimization, as this approach highlights the most significant impact on network's performance and resilience capability. Regarding the weights $\vect{\alpha}$, we employ a metric that weights the users based on their direct channel strengths, i.e., $\alpha_{1/2,k} = \frac{||\vect{h}_k||}{\max_{i\in\mathcal{K}}||\vect{h}_i||} \in [0,1]$.

\begin{figure}
  \centering
  \includegraphics[width=0.75\linewidth]{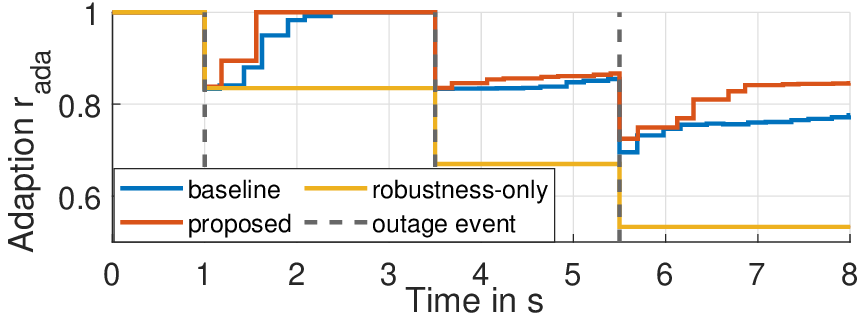}
  \caption{Adaption performance of the proposed gradient-augmentation approach, the baseline adaption gap-only and the robustness-only approach\vspace{-0.35cm}}
  \label{overallAda}
\end{figure}

\begin{figure}
  \centering
  \includegraphics[width=0.75\linewidth]{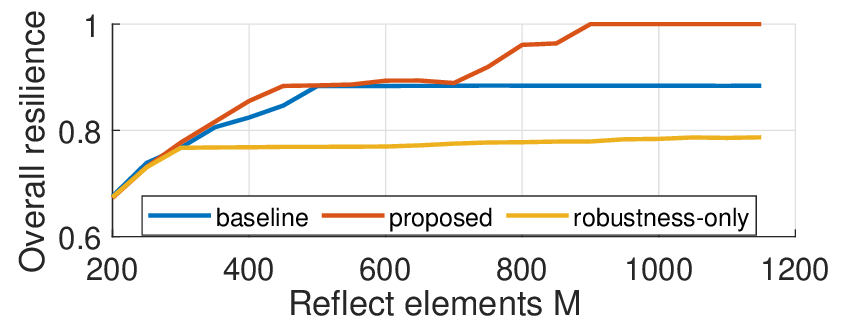}
  \caption{Overall resilience $r$ for the second blockage over the number of reflecting elements $M$ for the proposed, baseline and robustness-only approach\vspace{-1.5cm}}
  \label{overallR}
\end{figure}

Fig.~\ref{overallAda} displays the adaption performance $r_\mathsf{ada}$ for the proposed framework, which also maximizes the network's adaptability and compares it with the adaption performance of the baseline method proposed in \cite{RIS_RES}, which only optimizes the adaption gap $\Uppsi$. In addition, we include the robustness-only case in which $\alpha_{1,k}=0,\forall k\in\mathcal{K}$. The figure shows the significant increase in adaption performance of the proposed framework when compared to the baseline method. The results show that the proposed framework recovers from three consecutive blockages not only more quickly but also at a higher value than the other methods, utilizing the same resources. This demonstrates the effectiveness of gradient augmentation in the context of resilience. The figure also illustrates that prioritizing robustness alone can actually degrade absorption performance due to the resource overhead required to ensure full redundant network coverage via RIS-assisted paths only. This underscores the need to balance adaptability and robustness in resilience design, as shown in Fig.~\ref{overallR}, which presents the overall resilience (\ref{eq:r}) for the second blockage over different numbers of reflect elements $M$ with $\lambda_1=\lambda_2=\lambda_3=\frac{1}{3}$. The figure reveals distinct scaling behaviors among the three approaches. For $M<300$, resources are insufficient to meet user rate demands even pre-blockage. Beyond $M=300$, robustness-only scaling declines as using RIS paths only for redundancy becomes inefficient. In contrast, the adaption-based methods utilize resources more effectively, scaling until $M\approx500$, where they are able to fully recover from the first blockage. The proposed method with optimized adaptability achieves this earlier at $M=450$ due to faster resource reallocation enabled by gradient augmentation. At $M=700$, the proposed method begins to scale towards optimal resilience $r=1$, while the baseline stops scaling, struggling with reoptimization and suffering from absorption $r_\mathsf{abs}$ and time-to-recovery $r_\mathsf{rec}$ penalties.\vspace{-0.15cm}
\looseness-1
\looseness-1

\section{Conclusion}\label{ch:conc}
As wireless networks play a vital role in critical services, resilience becomes a crucial property, requiring not only robustness but also adaptability to react to unforeseen disruptions. Traditional approaches often neglect this aspect, limiting their ability to sustain performance in dynamic environments. To address this, we introduced a framework that explicitly captures adaptability and integrates \acp{RIS} to further improve resilience.
Results show that by utilizing gradient augmentation while continuously reallocating resources, our method maintains high resilience even after consecutive blockages, outperforming the baseline strategy that degrades more rapidly. This underscores the efficient use of available resources in a dynamic environment, laying the groundwork for uninterrupted network performance in future wireless systems.

\appendices
\section{Derivation of gradient $\nabla_\vect{v}r_k$}\label{sec:AppA}
Using the chain rule, we write the gradient of $r_k^\mathsf{RIS}$ in (\ref{RIS}) as
\begin{align}\label{eq:SINR_App}
\nabla_\vect{v} r_k^\mathsf{RIS} = \frac{\partial r_k^\mathsf{RIS}}{\partial \Gamma_k^\mathsf{RIS}}\nabla_\vect{v} \Gamma_k^\mathsf{RIS} = \frac{B}{\ln(2)}\frac{1}{1+\Gamma_k^\mathsf{RIS}}{\nabla_\vect{v} \Gamma_k^\mathsf{RIS}}.
\end{align} By denoting $\vect{h}_{k,i}^\vect{w} = \mat{G}_k^H \vect{w}_i$ and $\eta_{k,i} = \vect{v}^H \vect{h}_{k,i}^\vect{w}$, the RIS-only \ac{SINR} $\Gamma_k^\mathsf{RIS}$ can be written as
\begin{align}
\Gamma_k^\mathsf{RIS} =\frac{|\eta_{k,k}|^2}{\sum_{i \in\mathcal{K}\setminus \{k\} }|\eta_{k,i}|^2 + \sigma^2} = \frac{N_{\Gamma_k}}{D_{\Gamma_k}}.
\end{align} Applying the quotient rule, we arrive at
\begin{align}
&\nabla_\vect{v} \Gamma_k = \frac{ D_{\Gamma_k}  \nabla_\vect{v} N_{\Gamma_k} - N_{\Gamma_k}  \nabla_\vect{v} D_{\Gamma_k} }{D_{\Gamma_k}^2} = \frac{ \nabla_\vect{v} N_{\Gamma_k} - \Gamma_k  \nabla_\vect{v} D_{\Gamma_k} }{D_{\Gamma_k}}, \nonumber \\
&\nabla_\vect{v} N_{\Gamma_k}\hspace{-0.1cm}= \hspace{-0.1cm} 2\vect{h}_{k,k}^\vect{w}(\vect{h}_{k,k}^\vect{w})^H \vect{v}, \nabla_\vect{v} D_{\Gamma_k}\hspace{-0.1cm}= 2 \hspace{-0.33cm}\sum_{i\in\mathcal{K}\setminus{\{k}\}}\hspace{-0.3cm} \vect{h}_{k,i}^\vect{w}(\vect{h}_{k,i}^\vect{w})^H \vect{v}.\label{eq:quot}
\end{align}
By substituting (\ref{eq:quot}) into  (\ref{eq:SINR_App}) and simplifying $\frac{1}{1+\Gamma_k} \frac{1}{D_{\Gamma_k}} = \frac{1}{D_{\Gamma_k}+ N_{\Gamma_k}}$ we obtain
\begin{align}
   \nabla_\vect{v}r_k^\mathsf{RIS}= \frac{2B \big(\vect{h}_{k,k}^\vect{w}(\vect{h}_{k,k}^\vect{w})^H \vect{v} \hspace{-0.05cm} - \hspace{-0.05cm}  \Gamma_k \hspace{-0.2cm} \underset{i\in\mathcal{K}\setminus{\{k}\}}{\sum}\hspace{-0.2cm} \vect{h}_{k,i}^\vect{w}(\vect{h}_{k,i}^\vect{w})^H \vect{v}\big)}{\ln(2)(N_{\Gamma_k}+D_{\Gamma_k})}.
\end{align}
By recognizing that \((\vect{h}_{k,i}^\vect{w})^H \vect{v}\) is a scalar, we can transform \(\vect{h}_{k,k}^\vect{w}(\vect{h}_{k,k}^\vect{w})^H \vect{v}\) into \((\vect{h}_{k,k}^\vect{w})^H \vect{v}\,\vect{h}_{k,k}^\vect{w}\). This simplification facilitates further calculations in the paper, leading to equation (\ref{gradV}).

\balance
%\newpage
%\pagestyle{scrplain}
%\appendix
%
%\input{content/appendix}

%\cleardoublepage

%\cleardoublepage

%%

%\addcontentsline{toc}{chapter}{List of Figures}
%\listoffigures
%
%\listofalgorithms
%\cleardoublepage

%\addcontentsline{toc}{chapter}{List of Tables}
%\listoftables
%\cleardoublepage

%%
%\cleardoublepage

%\renewcommand*{\lstlistlistingname}{List of Listings}
%\lstlistoflistings
%\cleardoublepage

%\addcontentsline{toc}{chapter}{List of Formulas}
%\listofmyequations
%\cleardoublepage

%%

%\flushbottom
%\nocite*{}
% modified: /usr/share/texmf-texlive/bibtex/bst/dinat/dinat.bst

\footnotesize
\bibliographystyle{IEEEtran}
\bibliography{references}
\balance
%\section{Acronyms}
\begin{acronym}
\setlength{\itemsep}{0.1em}
\acro{AF}{amplify-and-forward}
\acro{AI}{artificial intelligence}
\acro{AP}{access point}
\acro{AWGN}{additive white Gaussian noise}
\acro{B5G}{Beyond 5G}
\acro{BS}{base station}
\acro{CB}{coherence block}
\acro{CE}{channel estimation}
\acro{C-RAN}{Cloud Radio Access Network}
\acro{CMD}{common message decoding}
\acro{CP}{central processor}
\acro{CSI}{channel state information}
\acro{CRLB}{Cramér-Rao lower bound}
\acro{D2D}{device-to-device}
\acro{DC}{difference-of-convex}
\acro{DFT}{discrete Fourier transformation}
\acro{DL}{downlink}
\acro{GDoF}{generalized degrees of freedom}
\acro{IC}{interference channel}
\acro{i.i.d.}{independent and identically distributed}
\acro{IRS}{intelligent reflecting surface}
\acro{IoT}{Internet of Things}
\acro{LoS}{line-of-sight}
\acro{LSF}{large scale fading}
\acro{KPI}{key performance indicator}
\acro{M2M}{Machine to Machine}
\acro{MISO}{multiple-input and single-output}
\acro{MIMO}{multiple-input and multiple-output}
\acro{MRT}{maximum ratio transmission}
\acro{MRC}{maximum ratio combining}
\acro{MSE}{mean square error}
\acro{NOMA}{non-orthogonal multiple access}
\acro{NLoS}{non-line-of-sight}
\acro{PSD}{positive semidefinite}
\acro{QCQP}{quadratically constrained quadratic programming}
\acro{QoS}{quality-of-service}
\acro{RF}{radio frequency}
\acro{RC}{reflect coefficient}
\acro{RIS}{reconfigurable intelligent surface}
\acro{RS-CMD}{rate splitting and common message decoding}
\acro{RSMA}{rate-splitting multiple access}
\acro{RS}{rate splitting}
\acro{SCA}{successive convex approximation}
\acro{SDP}{semidefinite programming}
\acro{SDR}{semidefinite relaxation}
\acro{SIC}{successive interference cancellation}
\acro{SINR}{signal-to-interference-plus-noise ratio}
\acro{SOCP}{second-order cone program}
\acro{SVD}{singular value decomposition }
\acro{TIN}{treating interference as noise}
\acro{TDD}{time-division duplexing}
\acro{TSM}{topological signal management}
\acro{UHDV}{Ultra High Definition Video}
\acro{UL}{uplink}
\acro{w.r.t.}{with respect to}

\acro{AF}{amplify-and-forward}
\acro{AWGN}{additive white Gaussian noise}
\acro{B5G}{Beyond 5G}
\acro{BS}{base station}
\acro{C-RAN}{Cloud Radio Access Network}
\acro{CSI}{channel state information}
\acro{CMD}{common-message-decoding}
\acro{CM}{common-message}
\acro{CoMP}{coordinated multi-point}
\acro{CP}{central processor}
\acro{D2D}{device-to-device}
\acro{DC}{difference-of-convex}
\acro{EE}{energy efficiency}
\acro{IC}{interference channel}
\acro{i.i.d.}{independent and identically distributed}
\acro{IRS}{intelligent reflecting surface}
\acro{IoT}{Internet of Things}
\acro{LoS}{line-of-sight}
\acro{LoSC}{level of supportive connectivity}
\acro{M2M}{Machine to Machine}
\acro{NOMA}{non-orthogonal multiple access}
\acro{MISO}{multiple-input and single-output}
\acro{MIMO}{multiple-input and multiple-output}
\acro{MMSE}{minimum mean squared error}
\acro{MRT}{maximum ratio transmission}
\acro{MRC}{maximum ratio combining}
\acro{NLoS}{non-line-of-sight}
\acro{PA}{power amplifier}
\acro{PSD}{positive semidefinite}
\acro{QCQP}{quadratically constrained quadratic programming}
\acro{QoS}{quality-of-service}
\acro{RF}{radio frequency}
\acro{RRU}{remote radio unit}
\acro{RS-CMD}{rate splitting and common message decoding}
\acro{RS}{rate splitting}
\acro{SDP}{semidefinite programming}
\acro{SDR}{semidefinite relaxation}
\acro{SIC}{successive interference cancellation}
\acro{SCA}{successive convex approximation}
\acro{SINR}{signal-to-interference-plus-noise ratio}
\acro{SOCP}{second-order cone program}
\acro{SVD}{singular value decomposition }
\acro{TP}{transition point}
\acro{TIN}{treating interference as noise}
\acro{UHDV}{Ultra High Definition Video}
\acro{LoSC}{level of supportive connectivity}
%\acro{M2M}{Machine to Machine}
%\acro{B5G}{Beyond 5G}
%\acro{CP}{Central Processor}
%\acro{IRS}{Intelligent Reflecting Surface}
%\acro{IoT}{Internet of Things}
%\acro{BS}{base station}
%\acro{C-RAN}{Cloud Radio Access Network}
%\acro{TIN}{treating interference as noise}
%\acro{RS}{rate splitting}
%\acro{CMD}{common message decoding}
%\acro{RS-CMD}{rate splitting and common message decoding}
%\acro{UHDV}{Ultra High Definition Video}
%\acro{LoS}{line-of-sight}
%\acro{NLoS}{non-line-of-sight}
%\acro{AF}{amplify-and-forward}
%\acro{RF}{radio frequency}
%\acro{QoS}{quality-of-service}
%\acro{QCQP}{quadratically constrained quadratic programming}
%\acro{DC}{difference-of-convex}
%\acro{IC}{interference channel}
%\acro{SIC}{Successive Interference Cancellation}
%\acro{AWGN}{Additive White Gaussian Noise}
%\acro{SINR}{signal-to-interference-plus-noise
%ratio}
%\acro{SINRs}{signal-to-interference-plus-noise
%ratios}
%\acro{MRC}{maximum ratio combining}
%\acro{D2D}{device-to-device}
%\acro{MIMO}{multiple-input and multiple-output}
%\acro{i.i.d.}{independent and identically distributed}
%\acro{SOCP}{second-order cone program}
%\acro{SDR}{semidefinite relaxation}
%\acro{SDP}{semidefinite programming}
%\acro{PSD}{positive semidefinite}
%\acro{SVD}{singular value decomposition}
\end{acronym}

\balance
\end{document}